\documentclass{article}
\usepackage{graphics,graphicx,psfrag,color,float, hyperref}
\usepackage{amsmath}
\usepackage{amssymb}

\usepackage{bm}

\usepackage{amssymb}
\usepackage{amsfonts}
\usepackage{amsthm, hyperref}
\usepackage {times}
\usepackage{amsfonts}

\newcommand{\beq}{\begin{equation}}
\newcommand{\enq}{\end{equation}}
\newcommand{\bel}{\begin{lemma}}
\newcommand{\enl}{\end{lemma}}
\newcommand{\bet}{\begin{theorem}}
\newcommand{\ent}{\end{theorem}}

\newcommand{\tr}{\mathrm{Tr}}

\newcommand{\tS}{\tilde{S}}
\newcommand{\ts}{\tilde{s}}
\newcommand{\eps}{\varepsilon}
\newcommand*{\cC}{\mathcal{C}}

\newcommand*{\cH}{\mathcal{H}}

\newcommand*{\cN}{\mathcal{N}}
\newcommand*{\cS}{\mathcal{S}}

\newcommand*{\cT}{\mathcal{T}}
\newcommand*{\cX}{\mathcal{X}}

\newcommand*{\cE}{\mathcal{E}}
\newcommand*{\cU}{\mathcal{U}}

\newcommand*{\cY}{\mathcal{Y}}

\newcommand*{\bU}{\mathbf{U}}

\newcommand{\bra}[1]{\langle #1|}
\newcommand{\ket}[1]{|#1 \rangle}

\newcommand*{\renyi}{R\'{e}nyi }

\newtheorem{definition}{Definition}

\newtheorem{corollary}{Corollary}

\newtheorem{theorem}{Theorem}
\newtheorem{lemma}{Lemma}

\begin {document}
\title{Coding for classical-quantum channels with rate limited side information at the encoder: An information-spectrum approach}
\author{Naqueeb Ahmad Warsi$^*$ 
\and 
Justin Coon
\thanks{
Department of Engineering Science, University of Oxford,
Email: 
{\sf 
 naqueeb.ahmedwarsi@eng.ox.ac.uk, justin.coon@eng.ox.ac.uk
}
}
}
\date{}

\maketitle

\begin{abstract}
We study the hybrid classical-quantum version of the channel coding problem for the famous Gel'fand-Pinsker channel. In the classical setting for this channel the conditional distribution of the channel output given the channel input is a function of a random parameter called the channel state. We study this problem when a rate limited version of the channel state is available at the encoder for the classical-quantum Gel'fand-Pinsker channel. We establish the capacity region for this problem in the information-spectrum setting. The capacity region is quantified in terms of spectral-sup classical mutual information rate and spectral-inf quantum mutual information rate.
\end{abstract}

\section{Introduction}
In traditional information theory literature it is common to study the
underlying problems assuming that the
channel characteristics do not change over multiple use. The proofs
appeal to {\em typicality} of sequences or typical subspaces in the quantum setting \cite{wilde-book}: the empirical
distribution of symbols in a long sequence of trials will with high
probability be close to the true
distribution~\cite{covertom}.  However, information
theoretic arguments based on typicality or the related Asymptotic
Equipartition Property (AEP) assume that both the source and channel
are stationary and/or ergodic (memoryless), assumptions that are not
always valid, for example, in \cite{gray-book} Gray analyzes the details of asymptotically mean stationary sources, which are neither stationary nor ergodic. To overcome such assumptions Verd\'{u} and Han pioneered the technique of information-spectrum methods in their seminal work \cite{han-verdu-spectrum-94}. In this work Verd\'{u} and Han define the notions of limit inferior and limit superior in probability. They then use these definitions to establish the  capacity of general channels (channels that are not necessarily stationary and/or memoryless). Since this work of Verd\'{u} and Han there have been a considerable interest in generalizing the results of information theory in the information spectrum setting, see for example,  \cite{miyakaye-kanaya-1995, muramatsu, hayashi-classical-non-iid, arbitrary-wiretap} and references therein.

This general technique of information-spectrum methods wherein no assumptions are made on the channels and/or sources were extended to the quantum case by Hayashi, Nagaoka  and Ogawa. Using this method they studied the problem of quantum hypothesis testing \cite{hayshi-nagaoka-2002, ogawa-nagaoka-2000-strong}, deriving the classical capacity formula of general quantum channels \cite{Hayashi-noniid} and establishing general formula for the optimal rate of entanglement concentration \cite{hayashi-entanglement-2006}. Since the work of Hayashi, Nagaoka and Ogawa the study of various quantum information theoretic protocols in the information spectrum setting have been one of the most interesting areas of research in the theoretical quantum information science. In \cite{datta-byeondiid-2006} Bowen and Datta further carried forward this approach to study various other quantum information theoretic problems. In \cite{datta-renner-2009} Datta and Renner showed that there is a close relationship with the information theoretic quantities that arise in the information-spectrum scenario and smooth \renyi entropies which play a crucial role in one-shot information theory. In \cite{radhakrishnan-sen-warsi-archive} Radhakrishnan, Sen and Warsi proved one-shot version of the Marton inner bound for the classical-quantum broadcast channels. They then showed that their one-shot bounds yields the quantum information-spectrum genralization of the Marton inner bound in the asymptotic setting. 

In this paper, we carry forward the subject of studying quantum information theoretic protocols in the information-spectrum setting. We study the problem of communication over the sequence $\left\{\cX^n,\cS^n, \cN^{X^nS^n \to B^n}(x^n,s^n) = \rho^{B^n}_{x^n,s^n}\right\}_{n=1}^\infty$ (also called as the classical-quantum Gel'fand-Pinsker channel), where $\cX^n,\cS^n$ are the input and state alphabets and $\rho^{B^n}_{x^n,s^n}$ is a positive operator with trace one acting on the Hilbert space $\cH_B^{\otimes n}.$ We establish the capacity region of this channel when rate limited version of the state sequence $S^n$ is available at the encoder. Figure \ref{Fig} below illustrates this communication scheme. 
\vspace{5mm}
\begin{figure}[H]
\centering
\includegraphics[scale=0.7] {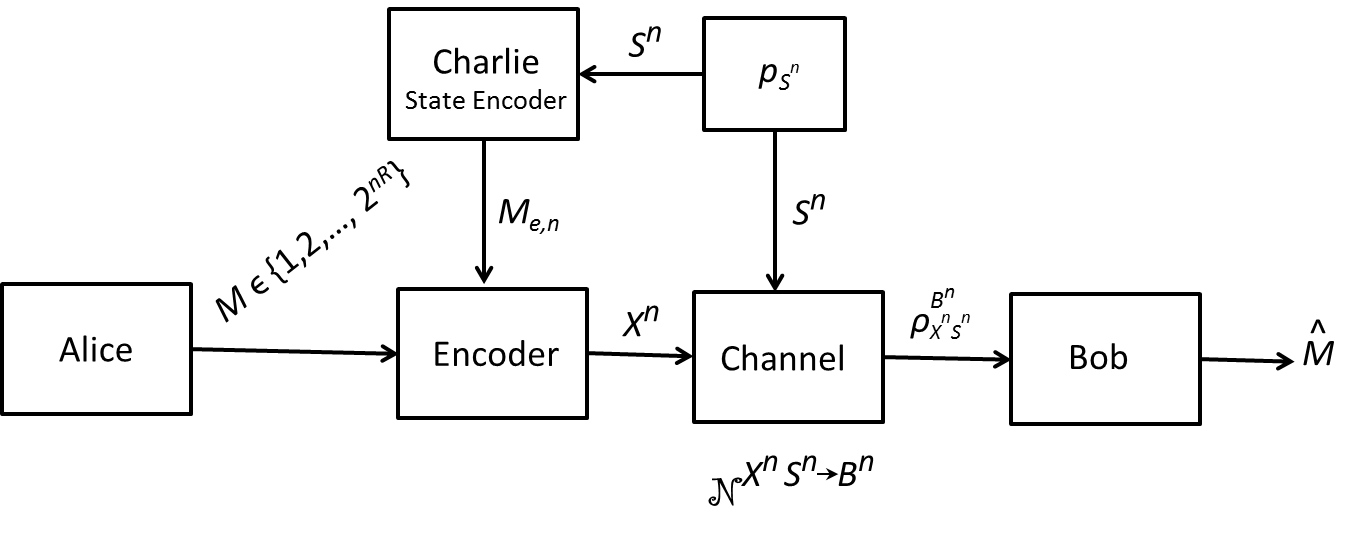}
\caption{Gel'fand-Pinsker channel communication scheme when coded side information is available at the encoder.}
\label{Fig}
\end{figure}

The classical version of this problem was studied by Heggard and El Gamal (achievability) in \cite{heegard-gamal-1983} in the asymptotic iid setting. They proved the following:
\begin{theorem}
\label{classical}
Fix  a discrete memoryless channel with state characterized by $p(y \mid x,s).$ Let $(R, R_S)$ be such that
\begin{align*}
R &< I[U;Y] - I [U; \tS]\\
R_ S&> I [S; \tS],
\end{align*}
for some distribution $p(s,\ts, u,x,y) = p(s)p(\ts \mid s)p(u \mid \ts) p(x \mid u, \ts)p(y \mid x,s).$ Then, the rate pair $(R,R_S)$ is achievable.
\end{theorem} 
Furthermore, in \cite{heegard-gamal-1983} Heggard and El Gamal argued that Theorem \ref{classical} implies the result of Gel'fand and Pinsker \cite{gelfand-pinsker} who showed the following:
\begin{theorem}
\label{gelf}
Fix  a discrete memoryless channel with state characterized by $p(y \mid x,s).$ The capacity of this channel when the state information is directly available non-causally at the encoder is
\beq
%\label{gelfc}
C = \max_{p_{u \mid s}, g : \cU \times \cS \to \cX} \left(I[U;Y] -I[U;S]\right), \nonumber
\enq
where $|\cU| \leq \min \left\{|\cX|.|\cS|, |\cY| + |\cS| -1\right\}.$
\end{theorem} 
The above formula for the capacity is quite intuitive. If we set $S =\emptyset$ and $U = X$ in Theorem \ref{gelf} then we rederive the famous Shannon's channel capacity formula \cite{shannon1948}. However, when $S \neq \emptyset$, Theorem \ref{gelf} implies that there is a loss in the maximum transmission rate per channel use at which Alice can communicate to Bob. This loss in the transmission rate is reflected by the term $I[U;S]$. Thus, $I[U;S]$ can be thought of as the minimum number of bits Alice needs to send to Bob per channel use to help him get some information about the channel state sequence $S^n$. Bob can then use this information about $S^n$ to recover the intended message.
\subsection*{Our Result}
We establish the capacity region of the classical-quantum Gel'fand-Pinsker channel in the information-spectrum setting when rate limited version of the channel state is available at the encoder. In the information-spectrum setting the channel output $\rho^{B^n}_{X^n,S^n},$ need not be a tensor product state. Furthermore, the channel state $S^n \sim p_{S^n},$ is a sequence of arbitrarily distributed random variables. This extremely general setting is the hallmark of information-spectrum approach.  We prove the following:
\begin{theorem}
\label{ourresult}
Let $\left\{\cX^n,\cS^n, \cN^{X^nS^n \to B^n}(x^n,s^n) = \rho^{B^n}_{x^n,s^n}\right\}_{n=1}^\infty$ be a sequence of classical-quantum Gel'fand-Pinsker channels. The capacity region for this sequence of channels with rate limited version of the channel state available only at the encoder is the set of rate pairs satisfying the following:
\begin{align*}
R &\leq \underline{\mathbf{I}}[{\bf{U}};{\bf{B}}] - \overline{\mathbf{I}}[\bf{U};\bf{\tilde{S}}];\\
R_S & \geq \overline{\mathbf{I}} [\bf{S};\bf{\tilde{S}}].
\end{align*}
The information theoretic quantities are calculated with respect to the sequence of states $\bm{\Theta^{S\tilde{S}UXB}} =\left\{\Theta^{S^n\tilde{S}^nU^nX^nB^n}\right\}_{n=1}^\infty,$ where for every $n$, 
\begin{align*}
&\Theta^{S^n\tilde{S}^nU^nX^nB^n}:= \sum_{s^n,\tilde{s}^n,u^,x^n}p_{{S^n}}({s}^n)p_{\tilde{S}^n \mid S^n}(\tilde{s}^n \mid s^n) p_{U^n \mid {\tilde{S}}^n}(u^n \mid \tilde{s}^n) p_{X^n \mid U^n\tS^n}(x^n \mid u^n,\ts^n)\ket{s^n}\bra{s^n}^{S^n}\otimes\ket{\tilde{s}^n}\bra{\tilde{s}^n}^{\tS^n}\\
&\hspace{45mm}\otimes\ket{u^n}\bra{u^n}^{U^n}\otimes\ket{x^n}\bra{x^n}^{X^n} \otimes \rho^{B^n}_{x^n,s^n}. \nonumber
\end{align*}
\end{theorem}
An immediate consequence of Theorem \ref{ourresult} is the following corollary:

\begin{corollary}
\begin{description}
\item[$(a)$] (Hayashi and Nagaoka, \cite{Hayashi-noniid}) The capacity of a sequence of classical-quantum channels $\{\cX^n, \cN^{X^n \to B^n}(x^n) = \rho^{B^n}_{x^n}\}_{n=1}^{\infty}$ is the following:
$$C = \sup_{\{X^n\}_{n=1} ^ \infty} \underline{\mathbf{I}} [\bf{X};\bf{B}].$$

\item [$(b)$] The capacity of a sequence of classical-quantum Gel'fand-Pinsker channels $\left\{\cX^n,\cS^n, \cN^{X^nS^n \to B^n}(x^n,s^n) = \rho^{B^n}_{x^n,s^n}\right\}_{n=1}^\infty$ with channel state directly available at the encoder is the following:
$$ C  = \sup_{\{\Theta_n\}_{n=1}^\infty}  \underline{\mathbf{I}} [\bf{U};\bf{B}] - \overline{\mathbf{I}} [\bf{U};\bf{S}],$$
where for every $n$, $$\Theta_n = \sum_{s^n,u^,x^n}p_{{S^n}}({s}^n)p_{U^n \mid {{S}}^n}(u^n \mid {s}^n) p_{X^n \mid U^nS^n}(x^n \mid u^n,s^n)\ket{s^n}\bra{s^n}^{S^n}\otimes\ket{u^n}\bra{u^n}^{U^n}\otimes\ket{x^n}\bra{x^n}^{X^n} \otimes \rho^{B^n}_{x^n,s^n}. $$

\end{description}
\end{corollary}
\begin{proof}
\begin{description}
\item[$a)$] The proof follows by setting $\tS^n=S^n = \emptyset$ and $U^n =X^n$ in Theorem \ref{ourresult}.
\item[$b)$] The proof follows by setting $\tS^n=\emptyset$ in Theorem \ref{ourresult}.
\end{description}

%\begin{description}
%{%\item[$a$)] %The achievability and converse follows by setting $\tS^n=S^n = \emptyset$ and $U^n =X^n$ in Theorem \ref{ourresult}}.
%{\item[$b$)] %The achievability and converse follows by setting $\tS^n=\emptyset$ in Theorem \ref{ourresult}}.
%\end{description}
\end{proof}

\section {Definition}

 \begin{definition}
 \label{limsup}
Let $(\bU,\mathbf{{\tS}}):=\left\{U^n,\tS^n\right\}_{n=1}^{\infty}$ be a sequence of pair of random variables, where for every $n$ $(U^n,\tS^n) \sim p_{U^nS^n}$ and take values over the set $(\cU^n \times \tilde{\mathcal{S}}^n)$. The spectral-sup mutual information rate $\overline{\mathbf{I}}[\bU;\mathbf{{\tS}}]$ between $\bU$ and $\mathbf{{\tS}}$ is defined as follows:
\beq
\overline{\mathbf{I}}[\bU;\mathbf{{\tS}}]:= \inf\left\{a: \lim_{n \to \infty }\Pr \left\{\frac{1}{n}\log\frac{p_{U^n\tS^n}}{p_{U^n} p_{\tS^n}} > a+\gamma\right\} = 0\right\},
\enq
where $\gamma >0$ is arbitrary and the probability above is calculated with respect to $p_{U^n\tS^n}$.
\end{definition}
 \begin{definition}
 Let ${\bm{\rho}}:= \left\{\rho_n\right\}_{n=1}^{\infty}$ and $\bm{\sigma}:=\left\{\sigma_n\right\}_{n=1}^\infty$ be sequences of quantum  states where for every $n,$ $\rho_n$ and $\sigma_n$ are density matrices acting on the Hilbert space $\cH_n:=\cH^{\otimes n}.$ The spectral-inf mutual information rate $\underline{\mathbf{I}}[\bm{\rho};\bm{\sigma}]$ between $\bm{\rho}$ and $\bm{\sigma}$ is defined as follows:
 \beq
 \underline{\mathbf{I}}[\bm{\rho};\bm{\sigma}]:=\sup \left\{a: \lim_{n\to \infty} \tr \left[\left\{\rho_n \succeq 2^{n(a-\gamma)} \sigma_n \right\} \rho_n\right] =1 \right\},
\enq
where $\gamma >0$ is arbitrary and $\left\{\rho_n \succeq 2^{n(a-\gamma)} \sigma_n \right\}$ represents a projection operator onto the non-negative Eigen space of the operator $\left(\rho_n - 2^{n(a-\gamma)} \sigma_n \right).$
 \end{definition}

\begin{definition}
\label{code}
An $(n,M_n,M_{e,n},\eps_n)$ code for the Gel'fand-Pinsker channel $\left\{\cX^n,\cS^n, \cN^{X^nS^n \to B^n}(x^n,s^n) = \rho^{B^n}_{x^n,s^n}\right\}$ with coded side information available at the encoder consists of 
\begin{itemize}
\item a state encoding $f_{e,n} : \cS^n \to [1:M_{e,n}]$
\item an encoding function $f_n: [1:M_n] \times [1:M_{e,n}] \to \cX^n$ (possibly randomized)
\item A decoding POVM $\left\{\beta(m): m \in [1:M_n]\right\}$ such that 
\beq
\frac{1}{M_n} \sum_{s^n}p_{S^n}(s^n) \tr\left[\left(\mathbb{I} - \cN(f_n(m, f_{e,n}(s^n)),s^n)\right)\beta(m)\right] \leq \eps_n. \nonumber
\enq
\end{itemize}
\end{definition}
\begin{definition}
\label{ach}
A rate pair $(R,R_S)$ is achievable if there exists a sequence of $(n,M_n,M_{e,n}, \eps_n)$ codes such that
\begin{align*}
\liminf _{n \to \infty} \frac{1}{n} \log M_n &>R\\
\limsup_{n \to \infty} \eps_n &< \eps \\
\limsup_{n \to \infty} \frac{1}{n} \log M_{e,n}&<R_S.
\end{align*} 
The set of all achievable rate pairs is known as the capacity region.
\end{definition}

\section{Proof of Theorem \ref{ourresult}}
\subsection{Achievability}
Let, 
\begin{align}
\label{rhofus}
\rho^{B^n}_{u^n,\ts^n}& = \sum_{s^n,x^n} p_{S^n \mid \tS^n}(s^n \mid \ts^n)p_{X^n \mid U^n\tS^n}(x^n \mid u^n,\ts^n)\rho^{B^n}_{x^n,s^n}\\
\label{theta}
\Theta^{U^nB^n}&= \tr_{S^n\tS^nX^n}\left[\Theta^{S^n\tS^nU^nX^nB^n}\right]\\
\Theta^{U^n} & = \tr_{B^n} \left[\Theta^{U^nB^n}\right]\\
\Theta^{B^n} & = \tr_{U^n} \left[\Theta^{U^nB^n}\right].
\end{align}
Let, 
\beq
\label{pi}
\Pi^{U^nB^n} := \left\{\Theta^{U^nB^n} \succeq 2^{n(\underline{\mathbf{I}}[{\bf{U}};{\bf{B}}] -\gamma)} \Theta^{U^n} \otimes \Theta^{B^n} \right\},
\enq
where $\underline{\mathbf{I}}[{\bf{U}};{\bf{B}}]$ is calculated with respect to the sequence of states $\left\{\Theta^{U^nB^n}\right\}_{n=1}^\infty$ and $\left\{\Theta^{U^n} \otimes \Theta^{B^n}\right\}_{n=1}^\infty.$ Further, for every $u^n \in \cU^n,$ let
\beq
\label{lambda}
\Lambda_{u^n} := \tr_{U^n}\left[\Pi^{U^nB^n}\left(\ket{u^n}\bra{u^n}\otimes \mathbb{I}\right)\right].
\enq

Fix $\gamma > 0.$ Define the following sets:
\begin{align*}
\cT_n(p_{S^n\tS^n})&:= \left\{(s^n,\ts^n) : \frac{1}{n}\log\frac{p_{S^n\tS^n}(s^n\ts^n)}{p_{S^n(s^n)}p_{\tS^n}(\ts^n)} \leq \overline{\mathbf{I}}[\bf{S};\bf{\tilde{S}}]+\gamma \right\};\\
\cT_n(p_{U^n\tS^n})&:= \left\{(u^n,\ts^n) : \frac{1}{n}\log\frac{p_{U^n\tS^n}(u^n\ts^n)}{p_{U^n(u^n)}p_{\tS^n}(\ts^n)} \leq \overline{\mathbf{I}}[\bf{U};\bf{\tilde{S}}]+\gamma\right\}.
\end{align*}
Furthermore, let $g_1: \tilde{\mathcal{S}}^n \to [0,1]$ and $g_2 : \tilde{\mathcal{S}}^n \to [0,1]$ be defined as follows: 
\begin{align}
\label{g}
g_1(\ts^n) = \sum_{u^n:(u^n,\ts^n) \notin \cT_n(p_{U^n\tS^n})}p_{U^n \mid \tS^n}(u^n \mid \ts^n);\\
\label{gg}
g_2(\ts^n) = \sum_{u^n:\tr \left[\Lambda_{u^n}\rho^{B^n}_{u^n, \ts^n}\right] \leq 1-\sqrt{\eps} } p_{U^n \mid \tS^n}(u^n \mid \ts^n).
\end{align} 
In what follows we will use the notation $[1:2^{nR}]$  to represent the set $\left\{1, \cdots, 2^{nR}\right\}.$

{\bf{The codebook:}}  Let $\Theta^{S^n\tilde{S}^nU^nX^nB^n}$ be as in the statement of the theorem. Let $R = \underline{\mathbf{I}}[{\bf{U}};{\bf{B}}] - \overline{\mathbf{I}}[\bf{U};\bf{\tilde{S}}]-6\gamma$, $r =  \overline{\mathbf{I}}[\bf{U};\bf{\tilde{S}}]+2\gamma$ so that $R+r = \underline{\mathbf{I}}[{\bf{U}};{\bf{B}}] -4\gamma$. Let $u^n[1],u^n[2],\cdots, u^n[2^{n(R+r)}]$ be drawn independently according to the distribution $p_{U^n}$. We associate these samples with a row vector $\cC_n^{(A)}$ having $2^{n(R+r)}$ entries. We then partition this row vector into $2^{nR}$ classes each containing $2^{nr}$ elements. Every message $m \in [1:2^{nR}]$ is uniquely assigned a class. We will denote the class corresponding to the message $m$ by $\cC_n^{(A)}(m)$.

Fix $R_S = \overline{\mathbf{I}} [\bf{S};\bf{\tilde{S}}] + 2\gamma .$ Further, let $\tilde{s}^n[1],\tilde{s}^n[2],\cdots,\tilde{s}^n[2^{nR_S}]$ be drawn independently according to the distribution $p_{\tS^n}$  We  will denote this collection of sequences by $\cC_n^{(C)}.$ These collection of sequences present in $\cC_n^{(C)}$ are made known to Alice as well.

{{\bf{Charlie's encoding strategy:}}}  For each $k \in [1:2^{nR_S}],$ let $Z(k)$ be independently and uniformly distributed over $[0,1].$ For a given realisation of the state sequence $s^n$, let $\zeta(k)$ be an indictor random variable defined as follows:
\beq
\label{jtyp11}
\mathbf{{\zeta}}(k)   =
\begin{cases}
1 & \mbox{if }~ Z(k) \leq
\frac{p_{\tS^nS^n}(\ts^n[k],s^n)}{2^{n(\overline{\mathbf{I}} [\bf{S};\bf{\tilde{S}}]+\gamma)}{ p_{\tS^n}(\ts^n[k]) p_{S^n}(s^n)}}; \\
0        & \mbox{otherwise.}
\end{cases}
\enq
Further, for a given realisation of the state sequence $s^n,$ let $\mathbf{I}(k)$ be an indicator random variable defined as follows:
\beq
\label{jtyp}
\mathbf{I}(k)   =
\begin{cases}
1 & \mbox{if }~
\zeta(k) =1 , g_1(\tS^n[k]) < \sqrt{\eps}~ \mbox{and}~ g_{2}(\tS^n[k]) < \eps^{\frac{1}{4}}; \\
0        & \mbox{otherwise,}
\end{cases}
\enq
where $g_1(\ts^n)$ and $g_2(\ts^n)$ are defined in \eqref{g} and \eqref{gg}. Charlie on observing the state sequence $s^n$ finds an index $k$ such that $\mathbf{I}(k)=1$. If there are more than one such indices then $k$ is set as the smallest one among them. If there is none such index then $k =1.$ Charlie then sends this index $k$ to Alice.

{\bf{Alice's encoding strategy:}}  For each pair $(k,\ell) \in [1:2^{nR_S}] \times [1:2^{n(R+r)}]$, let $\eta(k,l)$ be independently and uniformly distributed over $[0,1]$ and let $g(k,\ell)$ be defined as follows:
\beq
g(k,\ell):=\tr\left[\Lambda_{u^n[\ell]}\rho^{B^n}_{u^n[\ell],\ts^n[k]}\right],
\enq
where $\rho^{B^n}_{u^n,\ts^n}$ is defined in \eqref{rhofus} and $\Lambda_{u^n}$ is defined in \eqref{lambda}. Let $\mathbf{I}(k,\ell)$ be an indicator random variable such that 
\beq
\mathbf{I}(k,\ell)   =
\begin{cases}
1 & \mbox{if }~ \eta(k,\ell) \leq
\frac{p_{\tS^nU^n}(\ts^n[k],u^n[\ell])}{2^{n(\overline{\mathbf{I}} [\bf{U};\bf{\tilde{S}}]+\gamma)}{ p_{\tS^n}(\ts^n[k]) p_{U^n}(u^n[\ell])}}; \\
0        & \mbox{otherwise.}
\end{cases}
\enq
Further, let $\mathbf{J}(k,\ell)$ be an indicator random variable defined as follows:
\beq
\label{jtyp}
\mathbf{J}(k,\ell)   =
\begin{cases}
1 & \mbox{if }~ \mathbf{I}(k,\ell) =1~ \mbox{and} ~g(k,\ell) >1-\sqrt{\eps}; \\
0        & \mbox{otherwise.}
\end{cases}
\enq

To send a message $m \in [1:2^{nR}]$ and on receiving the index $k$ from Charlie, Alice finds an index $\ell\in \cC_n^{(A)}(m)$ such that $\mathbf{J}(k,\ell) =1.$ If there are more than such indices then $\ell$ is set as the smallest one among them. If there is none such index then 
$\ell =1.$ Alice then randomly generates $x^n \sim p_{X^n \mid U^n[\ell] \tS^n[k]}$ and transmits it over the classical-quantum channel over $n$ channel uses. In the discussions below we will use the notation $x^n(u^n[\ell], \ts^n[k])$ to highlight the dependence of $x^n$ on $(u^n[k], \ts^n[k])$. A similar encoding technique was also used by Radhakrishnan, Sen and Warsi in \cite{radhakrishnan-sen-warsi-archive}.

{\bf{Bobs' decoding strategy:}} For each $\ell \in [1:2^{n(R+r)}],$ we have the operators $\Lambda_{u^n[\ell]}$ as defined in \eqref{lambda}. Bob will normalize these operators to obtain a POVM. The POVM element corresponding to $\ell$ will be
\beq
\beta_n(\ell) := \left(\sum_{ \ell^\prime \in [1:2^{n(R+r)}]} \Lambda_{u^n(\ell^\prime)}\right)^{-\frac{1}{2}}\Lambda_{u^n(\ell)}\left(\sum_{ \ell^\prime \in [1:2^{n(R+r)}]} \Lambda_{u^n(\ell^\prime)}\right)^{-\frac{1}{2}}.
\enq
Bob on receiving the channel output measures it using these operators. If the measurement outcome is $\tilde{\ell}$ then he outputs $\tilde{m}$ if $\tilde{\ell} \in \cC_n^{(A)}(\tilde{m}).$ Similar decoding POVM elements were also used by Wang and Renner in \cite{wang-renner-prl}.

{\bf{Probability of error analysis:}}  Let a message $m \in [1:2^{nR}]$ be transmitted by Alice by using the protocol discussed above and suppose it is decoded as $\tilde{m}$ by Charlie. We will now show that the probability $\tilde{m} \neq m,$ averaged over the random choice of codebook, the state sequence $S^n$ and $X^n$ is arbitrary close to zero. By the symmetry of the code construction it is enough to prove the claim for $m=1.$ There are following sources of error:
 \begin{enumerate} 
 \item Charlie on observing the state sequence $S^n$ does not find a suitable $k \in \cC_n^{(C)}$ such that $\mathbf{I}(k)=1$.
 \item Alice on receiving the index $k$ from Charlie is not able to find a suitable $\ell \in \cC_n^{(A)}(1)$ such that $(U^n[\ell],\tilde{S}^n[k])$ such that $\mathbf{J}(k,\ell)=1.$
 \item Charlie finds a suitable $k$ and Alice finds a suitable $\ell$, but Bob's measurement is not able to determine the index $\ell$ correctly.
 \end{enumerate}
Let $k^\star$ and $\ell^\star$ be the indices chosen by Charlie and Alice. Let us now upper bound the probability of error while decoding the transmitted message. Towards this we first define the following events:
\begin{align*}
\cE_1&:= \mbox{for all}~k \in [1:2^{nR_s}] : \mbox{I}(k)=0;\\
\cE_2& := \mbox{for all}~\ell \in \cC_n^{(A)}(1) : \mbox{J}(k^\star, \ell) = 0.\\
%cE_3 & := \cE_1^c ~\mbox{and}~ \cE_2.\\
%cE_4 & := \cE_2^c ~\mbox{and} ~ \tilde{\ell} \neq \ell^\star
\end{align*}
We now have the following bound on the error probability: 
\begin{align}
\Pr\left\{\tilde{m} \neq 1 \right\} &\leq \Pr \left\{ \tilde{\ell} \neq \ell^\star\right\}\nonumber\\
& \leq \Pr\left\{\cE_1 \cup \cE_2 \right\} + \Pr \left\{ \left(\cE_1\cup \cE_2\right)^c,  \tilde{\ell} \neq \ell^\star\right\}\nonumber\\
& \leq \Pr\left\{\cE_1\right\} + \Pr\left\{\cE_2\right\} + \Pr \left\{\cE_1^c \cap \cE^c_2, \tilde{\ell} \neq \ell^\star\right\}  \nonumber\\
%& \leq \Pr\left\{\cE_1\right\} + \Pr\left\{\cE_1^c, \cE_2\right\} +\Pr \left\{\cE^c_2, \tilde{\ell} \neq \ell^\star\right\} \nonumber\\
\label{errorsideinf}
& \leq 2\Pr\left\{\cE_1\right\} + \Pr\left\{\cE_1^c \cap \cE_2\right\} +\Pr \left\{\cE_1^c \cap \cE^c_2, \tilde{\ell} \neq \ell^\star\right\}, 
\end{align}
where the first inequality follows from the setting of the protocol discussed above and remaining all of the inequalities till \eqref{errorsideinf} follow from the union bound. In what follows we will now show that for $n$ large enough we have
$$ 
2\Pr\left\{\cE_1\right\} + \Pr\left\{\cE_1^c \cap \cE_2\right\} +\Pr \left\{\cE_1^c \cap \cE^c_2, \tilde{\ell} \neq \ell^\star\right\} \leq 6\eps+3\sqrt{\eps} + 3\eps^{\frac{1}{4}}+\frac{2\sqrt{\eps}}{\left(1- \eps -\sqrt{\eps} -\eps^{\frac{1}{4}}\right)}+3\exp(-2^{n\gamma}), \nonumber
$$
where $\eps>0$ is arbitrarily close to zero such that $\left(\eps + \sqrt{\eps} + \eps^{\frac{1}{4}}\right) <1 $.

{\bf{Consider $\Pr \left\{\cE_1\right\}:$}} 
\begin{align}
\Pr\left\{\cE_1\right\} & \overset{a} = \sum_{s^n \in \cS^n}p_{S^n}(s^n) \left(1- \Pr\left\{\mathbf{I}(k)=1 \mid S^n =s^n\right\}\right)^{2^{nR_S}}\nonumber\\
 &\leq \sum_{s^n \in \cS^n}p_{S^n}(s^n) \left(1- \sum_{\substack{\ts^n:(\ts^n,s^n) \in \cT_n(p_{S^n\tS^n})\\ g_1(\ts^n) < \sqrt{\eps} \\ g_2(\ts^n) <\eps^{\frac{1}{4}} }}p_{\tS^n}(\ts^n)\Pr\left\{\mathbf{I}(k)=1 \mid S^n =s^n,\tS^n[k] =\ts^n\right\}\right)^{2^{nR_S}} \nonumber\\
& \overset{b} = \sum_{s^n \in \cS^n}p_{S^n}(s^n)\left(1-  2^{-n(\overline{\mathbf{I}}[\bf{S};\bf{\tilde{S}}]+\gamma)}\sum_{\substack{ \ts^n:(\ts^n,s^n) \in \cT_n(p_{S^n\tS^n})\\g_1(\ts^n) < \sqrt{\eps} \\ g_2(\ts^n) < \eps^{\frac{1}{4}}}}p_{\tS^n}(\ts^n)\frac{p_{S^n\tS^n}(s^n, \ts^n)}{p_{S^n}(s^n)p_{\tS^n}(s^n)}\right)^{2^{nR_S}} \nonumber\\ 
&\overset{c} \leq \sum_{s^n \in \cS^n}p_{S^n}(s^n)\exp\left(-2^{n(R_S-(\overline{\mathbf{I}} [\bf{S};\bf{\tilde{S}}] + \gamma))}\sum_{\substack{\ts^n: (\ts^n,s^n) \in \cT_n(p_{S^n\tS^n})\\ g_1(\ts^n) < \sqrt{\eps} \\ g_2(\ts^n) < \eps^{\frac{1}{4}}}}p_{\tS^n \mid S^n} (\ts^n \mid s^n)    \right) \nonumber\\
&\overset{d} = \sum_{s^n \in \cS^n}p_{S^n}(s^n)\exp\left(-2^{n\gamma}\sum_{\substack{\ts^n:(\ts^n,s^n) \in \cT_n(p_{S^n\tS^n})\\g_1(\ts^n) < \sqrt{\eps} \\ g_2(\ts^n) < \eps^{\frac{1}{4}}}} p_{\tS^n \mid S^n} (\ts^n \mid s^n)    \right) \nonumber\\
\label{markov}
& \overset{e} \leq \Pr\left\{\cT^c_n(p_{S^n\tS^n})\right\} + \Pr\left\{g_1(\tS^n) \geq \sqrt{\eps}\right\} + \Pr\left\{g_2(\tS^n) \geq \eps^{\frac{1}{4}}\right\} + \exp(-2^{n\gamma}) \nonumber\\
& \overset{f} \leq \eps + \exp(-2^{n\gamma}) + \Pr\left\{g_1(\tS^n) \geq \sqrt{\eps}\right\} + \Pr\left\{g_2(\tS^n) \geq \eps^{\frac{1}{4}}\right\},
\end{align}
where $a$ follows because $\mathbf{I}(1),\cdots,\mathbf{I}(2^{nR_S})$ are independent and identically distributed and $\tS^n[1], \cdots, \tS^n[2^{nR_S}]$ are independent and identically distributed according to the distribution $p_{\tS^n},$ $b$ follows from the definition of $\mathbf{I}(k),$ $c$ follows from the inequality $(1-x)^y \leq e^{-xy}(0\leq x\leq1, y\geq 0),$ $d$ follows becuase $R_s=\overline{{\mathbf{I}}}[\bf{U};\bf{\tilde{S}}]+2\gamma,$ $e$ follows because $(e^{-xy}) \leq 1-x+e^{-y} (0\leq x\leq1, y\geq 0)$ and union bound and $f$ follows because $n$ is large enough such that $\Pr\left\{\cT^c_n(p_{S^n\tS^n})\right\} \leq \eps.$ Let us now bound the each of the last two terms on the R.H.S. of \eqref{markov} as follows:

{\bf{Consider $\Pr\left\{g_1(\tS^n) \geq \sqrt{\eps}\right\}$}:}
\begin{align}
\Pr\left\{g_1(\tS^n) \geq \sqrt{\eps} \right\} & \overset{a} \leq \frac{\mathbb{E}[g_1(\tS^n)]}{\sqrt{\eps}} \nonumber\\
& \overset{b}= \frac{\sum_{\ts^n}\sum_{u^n:(u^n,\ts^n) \notin \cT_n(p_{U^n\tS^n})}p_{U^n \tS^n}(u^n,\tS^n)}{\sqrt{\eps}}\nonumber\\
& = \frac{\Pr\left\{(U^n,\tS^n) \notin \cT_n(p_{U^n\tS^n})\right\}}{\sqrt{\eps}}\nonumber\\
& \overset{c} \leq \frac{\eps}{\sqrt{\eps}} \nonumber\\
\label{g1m}
& =\sqrt{\eps},
\end{align}
where $a$ follows from Markov inequality; $b$ follows from the definition of $g_{1}(\ts^n)$ and taking expectation over the random variable $\tS^n$ and $c$ follows under the assumption that $n$ is large enough such that $\Pr\left\{(U^n,\tS^n) \notin \cT_n(p_{U^n\tS^n})\right\} \leq \eps.$

{\bf{Consider $\Pr\left\{g_2(\tS^n) \geq {\eps}^{\frac{1}{4}}\right\}$}:}
\begin{align}
\Pr\left\{g_2(\tS^n) \geq {\eps}^{\frac{1}{4}} \right\} & \overset{a} \leq \frac{\mathbb{E}[g_2(\tS^n)]}{{\eps}^{\frac{1}{4}}} \nonumber\\
& \overset{b} =\frac{\sum_{\ts^n}\sum_{u^n:\tr \left[\Lambda_{u^n}\rho^{B^n}_{u^n, \ts^n}\right] \leq 1-\sqrt{\eps} } p_{U^n\tS^n}(u^n,\ts^n)}{\eps^{\frac{1}{4}}} \nonumber\\
& = \frac{\Pr\left\{\tr \left[\Lambda_{U^n}\rho^{B^n}_{U^n, \tS^n}\right] \leq 1-\sqrt{\eps}\right\}}{\eps^{\frac{1}{4}}} \nonumber\\
&\overset{c} \leq \frac{\sqrt{\eps}}{\eps^{\frac{1}{4}}} \nonumber\\
\label{g2m}
& = \eps^{\frac{1}{4}},
\end{align}
where $a$ follows from Markov inequality; $b$ follows from the definition of $g_{2}(\ts^n)$ and by taking the expectation over the random variable $\tS^n$ and $c$ follows because of the following set of inequalities: 

\begin{align}
&\Pr\left\{\tr \left[\Lambda_{U^n}\rho^{B^n}_{U^n, \tS^n}\right] \leq1-\sqrt{\eps}\right\} \nonumber\\
 &\overset{a}
\leq \frac{1- \mathbb{E} \tr\left[\Lambda_{U^n}\rho^{B^n}_{U^n,\tS^n}\right]}{\sqrt{\eps}}\nonumber\\
& \overset{b}=\frac{1- \mathbb{E} \tr\left[\tr_{U^n}\left[\Pi^{U^nB^n}\left(\ket{U^n}\bra{U^n}\otimes \mathbb{I}\right)\right]\rho^{B^n}_{U^n,\tS^n}\right]}{\sqrt{\eps}}\nonumber \\
& = \frac{1- \tr\left[\tr_{U^n}\left[\Pi^{U^nB^n}\sum_{u^n}p_{U^n}(u^n)\left(\ket{u^n}\bra{u^n}\otimes \sum_{\ts^n}p_{\tS^n \mid U^n}(\ts^n \mid u^n)\rho^{B^n}_{u^n,\ts^n}\right)\right]\right]}{\sqrt{\eps}}\nonumber \\
& \overset{c}=   \frac{1- \tr \left[\Pi^{U^nB^n}\Theta^{U^nB^n}\right]}{\sqrt{\eps}}\nonumber\\
& \overset{d} \leq   \sqrt{\eps},\nonumber
\end{align} 
where $a$ follows from the Markov inequality, $b$ follows from the definition of $\Lambda_{U^n}$ mentioned in \eqref{lambda}, $c$ follows from the definition of $\Theta^{U^nB^n}$ mentioned in \eqref{theta} and $d$ follows under the assumption that $n$ is large enough such that $\tr \left[\Pi^{U^nB^n}\Theta^{U^nB^n}\right] \geq 1-\eps.$ Thus, it now follows from \eqref{markov}, \eqref{g1m} and \eqref{g2m} that
\beq
\label{e1}
\Pr\left\{\cE_1\right\} \leq \eps+ \sqrt{\eps}+ \eps^{\frac{1}{4}} + \exp(-2^{n\gamma}).
\enq

{\bf{Consider $\Pr\left\{\cE_1^c \cap \cE_2\right\}:$}} 
\begin{align}
&\Pr\left\{\cE_1^c \cap \cE_2\right\} \nonumber \\
& = \mathbb{E}_{\cC_n^{(C)} S^n}\mathbf{I}(\cE^c_1) \Pr \left\{\forall \ell \in \cC(1) : \mathbf{J}(k^\star,\ell)=0\right\} \nonumber\\
&\overset{a} =  \mathbb{E}_{\cC_n^{(C)}S^n}\mathbf{I}(\cE^c_1) \left(1- \Pr\left\{\mathbf{J}(k^\star,\ell)=1\right\}\right)^{2^{nr}} \nonumber\\
&\overset{b} = \mathbb{E}_{\cC_n^{(C)} S^n}\mathbf{I}(\cE^c_1) \left(1- \Pr\left\{\mathbf{I}(k^\star,\ell) =1, g(k^\star,\ell) > 1- \sqrt{\eps}\right\}\right)^{2^{nr}} \nonumber\\
& \leq  \mathbb{E}_{\cC_n^{(C)} S^n}\mathbf{I}(\cE^c_1) \left(1-  \sum_{\substack{u^n: (\tS^n[k^\star], u^n) \in \cT_{n}(p_{U^n\tS^n})\\ \tr \left[\Lambda_{u^n}\rho^{B^n}_{u^n, \tS^n[k^\star]}\right] >1-\sqrt{\eps}}}2^{-n(\overline{\mathbf{I}}[\bf{U};\bf{\tilde{S}}]+\gamma)}p_{U^n}(u^n)\frac{p_{\tS^nU^n}(\tS^n[k^\star],u^n)}{p_{\tS^n}(\tS^n[k^\star])p_{U^n}(u^n)} \right)^{2^{nr}}\nonumber\\
&\overset{c} \leq  \mathbb{E}_{\cC_n^{(C)} S^n}\mathbf{I}(\cE^c_1)\exp\left(-2^{n(r-(\overline{\mathbf{I}}[\bf{U};\bf{\tilde{S}}])+\gamma))} \sum_{\substack{u^n: (\tS^n[k^\star], u^n) \in \cT_{n}(p_{U^n\tS^n})\\ \tr \left[\Lambda_{u^n}\rho^{B^n}_{u^n, \tS^n[k^\star]}\right] >1-\sqrt{\eps}}}p_{U^n \mid \tS^n}(u^n\mid \tS^n[k^\star])\right)\nonumber\\
& \overset{d}\leq \mathbb{E}_{\cC_n^{(C)} S^n} \mathbf{I}(\cE^c_1)\exp\left(-2^{n\gamma} \sum_{\substack{u^n: (\tS^n[k^\star], u^n) \in \cT_{n}(p_{U^n\tS^n})\\ \tr \left[\Lambda_{u^n}\rho^{B^n}_{u^n, \tS^n[k^\star]}\right] >1-\sqrt{\eps}}}p_{U^n \mid \tS^n}(u^n\mid \tS^n[k^\star])\right)\nonumber\\
&\overset{e}\leq  \mathbb{E}_{\cC_n^{(C)}S^n}\mathbf{I}(\cE^c_1)\left(1- \sum_{\substack{u^n: (\tS^n[k^\star], u^n) \in \cT_{n}(p_{U^n\tS^n})\\ \tr \left[\Lambda_{u^n}\rho^{B^n}_{u^n, \tS^n[k^\star]}\right] >1-\sqrt{\eps}}}p_{U^n \mid \tS^n}(u^n\mid \tS^n[k^\star])\right)+
 \mathbb{E}_{\cC_n^{(C)} S^n}\mathbf{I}(\cE^c_1)\exp(-2^{n\gamma}) \nonumber\\
&\overset{f}\leq  \mathbb{E}_{\cC_n^{(C)}S^n}\mathbf{I}(\cE^c_1) \sum_{u^n: (\tS^n[k^\star], u^n) \notin\cT_{n}(p_{U^n\tS^n})}p_{U^n \mid \tS^n}\left(u^n\mid \tS^n[k^\star]\right) + \mathbb{E}_{\cC_n^{(C)} S^n}\mathbf{I}(\cE^c_1)\exp(-2^{n\gamma})\nonumber\\
&\hspace{5mm}+ \mathbb{E}_{\cC_n^{(C)} S^n}\mathbf{I}(\cE^c_1) \sum_{u^n: \tr \left[\Lambda_{u^n}\rho^{B^n}_{u^n, \tS^n[k^\star]}\right] \leq1-\sqrt{\eps}}p_{U^n \mid \tS^n}\left(u^n\mid \tS^n[k^\star]\right)\nonumber\\
\label{n1}
&\overset{g}\leq \sqrt{\eps} + \eps^{\frac{1}{4}} + \exp(2^{-n\gamma}),
\end{align}
where $a$ follows because $\eta(k^\star,1), \cdots \eta(k^\star,2^{nr})$ are independent and identically distributed and the fact that $U^n[1], \cdots,U^n[2^{nr}]$ are independent and subject to identical distribution $p_{U^n}$, $b$ follows from the definition of $\mathbf{J}(k^\star,\ell)$; $c$ follows from the fact that $(1-x)^y \leq e^{-xy} (0\leq x \leq 1, y \geq 0)$, $d$ follows because $r = \overline{\mathbf{I}} [\bf{U};\bf{\tilde{S}}]+2\gamma,$ $e$ follows because $ e^{-xy} \leq 1- x + e^{-y} (0\leq x \leq 1, y \geq 0)$; $f$ follows because of the union bound and $g$ follows because if the event $\cE^c_1$ happens then $\sum_{u^n: (\tS^n[k^\star], u^n) \notin\cT_{n}(p_{U^n\tS^n})}p_{U^n \mid \tS^n}\left(u^n\mid \tS^n[k^\star]\right) < \sqrt{\eps}$ and $\sum_{u^n: \tr \left[\Lambda_{u^n}\rho^{B^n}_{u^n, \tS^n[k^\star]}\right] \leq1-\sqrt{\eps}}p_{U^n \mid \tS^n}\left(u^n\mid \tS^n[k^\star]\right) < \eps^{\frac{1}{4}}.$ 

{\bf{Consider {$\Pr \left\{\cE_1^c \cap \cE^c_2, \tilde{\ell} \neq \ell^\star\right\}$:}}} 
\begin{align}
\Pr \left\{\cE_1^c \cap \cE^c_2, \tilde{\ell} \neq \ell^\star\right\} &=\mathbb{E}_{\cC_n^{(A)}\cC_n^{(C)}X^nS^n} \left[\mathbf{I}\left\{\cE_1^c\right\}\mathbf{I}\left\{\cE^c_2\right\} \tr\left[\left(\mathbb{I}-\beta_n(\ell^\star)\right) \rho^{B^n}_{X^n(U^n[\ell^\star],\tilde{S}^n[k^{\star}]),S^n}\right]\right]\nonumber\\
& \leq 2 \mathbb{E}_{\cC_n^{(A)}\cC_n^{(C)}X^nS^n}\left[\mathbf{I}\left\{\cE_1^c\right\}\mathbf{I}\left\{\cE^c_2\right\} \tr\left[\left(\mathbb{I}-\Lambda_{U^n[\ell^\star]}\right) \rho^{B^n}_{X^n(U^n[\ell^\star],\tilde{S}^n[k^{\star}]),S^n}\right] \right] \nonumber\\
\label{firster}
&\hspace{4mm}+ 4 \sum_{\ell^\prime \neq \ell^\star}\mathbb{E}_{\cC_n^{(A)}\cC_n^{(C)}X^nS^n}\left[\mathbf{I}\left\{\cE_1^c\right\}\mathbf{I}\left\{\cE^c_2\right\} \tr\left[\Lambda_{U^n(\ell^\prime)} \rho^{B^n}_{X^n(U^n[\ell^\star],\tilde{S}^n[k^{\star}]),S^n}\right] \right],
\end{align}
where the inequality above follows from the Hayashi-Nagaoka operator inequality \cite{Hayashi-noniid} .
In what follows we show that for $n$ large enough,
\begin{align}
\label{nontyp}
 2\mathbb{E}_{\cC_n^{(A)}\cC_n^{(C)}X^nS^n}\left[\mathbf{I}\left\{\cE_1^c\right\}\mathbf{I}\left\{\cE^c_2\right\} \tr\left[\left(\mathbb{I}-\Lambda_{U^n[\ell^\star]}\right) \rho^{B^n}_{X^n(U^n[\ell^\star],\tilde{S}^n[k^{\star}]),S^n}\right] \right] \leq \frac{2\sqrt{\eps}}{\left(1- \eps - \sqrt{\eps} -\eps^{\frac{1}{4}}\right)},
\end{align}
and 
\begin{align}
\label{othertyp}
4 \sum_{\ell^\prime \neq \ell^\star}\mathbb{E}_{\cC_n^{(A)}\cC_n^{(C)}X^nS^n}\left[\mathbf{I}\left\{\cE_1^c\right\}\mathbf{I}\left\{\cE^c_2\right\} \tr\left[\Lambda_{U^n(\ell^\prime)} \rho^{B^n}_{X^n(U^n[\ell^\star],\tilde{S}^n[k^{\star}]),S^n}\right] \right] \leq 4\eps.
\end{align}
We would like to highlight here that the proof pertaining to the derivation of \eqref{nontyp} is nontrivial and requires careful analysis of the probabilistic terms involved. In fact, the proof for the derivation of \eqref{othertyp} is nontrivial as well. However, we would borrow the idea of \emph{over-counting} from \cite{radhakrishnan-sen-warsi-archive} to bound \eqref{othertyp}. The reason for all this nontrivallity is because $k^\star$ and $\ell^\star$ are random. 

{{$2\mathbb{E}_{\cC_n^{(A)}\cC_n^{(C)}X^nS^n}\left[\mathbf{I}\left\{\cE_1^c\right\}\mathbf{I}\left\{\cE^c_2\right\} \tr\left[\left(\mathbb{I}-\Lambda_{U^n[\ell^\star]}\right) \rho^{B^n}_{X^n(U^n[\ell^\star],\tilde{S}^n[k^{\star}]),S^n}\right] \right]$ is bounded as follows:}}
\begin{align}
&2\mathbb{E}_{\cC_n^{(A)}\cC_n^{(C)} X^n S^n}\left[\mathbf{I}\{\cE_1^c\}\mathbf{I}\left\{\cE^c_2\right\} \tr\left[\left(\mathbb{I}-\Lambda_{U^n[\ell^\star]}\right) \rho^{B^n}_{X^n(U^n[\ell^\star],\tilde{S}^n[k^{\star}]),S^n}\right] \right] \nonumber\\
& \hspace{-1mm}= 2\sum_{u^n,\tilde{s}^n,x^n,s^n} \Pr\bigg\{\mathbf{I}(k^\star)=1,\mathbf{J}(k^\star,\ell^\star)=1, U^n[\ell^\star]=u^n, \tilde{S}^n[k^\star] = \tilde{s}^n, \nonumber\\
&\hspace{30mm} X^n(U^n[\ell^\star] =u^n, \tS^n [k^\star] =\ts^n) = x^n, S^n=s^n\bigg\} \nonumber\\
&\hspace{30mm} \tr\left[\left(\mathbb{I}-\Lambda_{(U^n[\ell^\star] = u^n)} \right) \rho^{B^n}_{X^n(U^n[\ell^\star] = u^n,\tS^n[k^\star]=\tilde{s}^n),S^n = s^n}\right]\nonumber\\
&\hspace{-1mm} \leq2\sum_{\ts^n,s^n,u^n,x^n} \Pr\left\{\tS^n[k^\star]=\ts^n, S^n=s^n\mid \mathbf{I}(k^\star)=1\right\} \Pr\bigg\{ U^n[\ell^\star]=u^n\mid S^n = s^n, \tS^n[k^\star]= \tS^n, \nonumber\\
 &\hspace{30mm}\mathbf{I}(k^\star) =1, \mathbf{J}(k^\star,\ell^\star)=1\bigg\}
\Pr\left\{X^n = x^n \mid U^n[\ell^\star]=u^n, \tS^n[k^\star]=\ts^n, S^n= s^n\right\} \nonumber\\
\label{ne}
&\hspace{30mm}\tr\left[\left(\mathbb{I}-\Lambda_{(U^n[\ell^\star]=u^n)}\right) \rho^{B^n}_{X^n(U^n[\ell^\star]=u^n,\tS^n[k^\star] = \tilde{s}^n) =x^n, S^n = s^n}\right], 
\end{align}
where the above inequality follows because $X^n$ given $(U^n[\ell^\star]=u^n, \tS^n[k^\star]=\ts^n, S^n= s^n)$ is conditionally independent of $(\mathbf{I}(k^\star), \mathbf{J}(k^\star,\ell^\star)).$ To get the required bound on $2\mathbb{E}_{\cC_n^{(A)}\cC_n^{(C)} X^n S^n}\left[\mathbf{I}\{\cE_1^c\}\mathbf{I}\left\{\cE^c_2\right\} \tr\left[\left(\mathbb{I}-\Lambda_{U^n[\ell^\star]}\right) \rho^{B^n}_{X^n(U^n[\ell^\star],\tilde{S}^n[k^{\star}]),S^n}\right] \right],$ we will now first show that $\Pr\left\{\tS^n[k^\star]=\ts^n, S^n=s^n\mid \mathbf{I}(k^\star)=1\right\} \leq \frac{p_{\tS^n S^n}(\ts^n ,s^n)}{\left(1- \eps-  \sqrt{\eps} -\eps^{\frac{1}{4}}\right)}.$  Towards this notice the following set of inequalities: 
\begin{align}
&\Pr\left\{\tS^n[k^\star]=\ts^n, S^n=s^n  \mid \mathbf{I}(k^\star)=1\right\}\nonumber\\
& = \sum_{k}\Pr\left\{k^\star= k \mid \mathbf{I}(k^\star)=1 \right\}\Pr\left\{\tS^n[k^\star]=\ts^n, S^n=s^n  \mid \mathbf{I}(k^\star)=1, k^\star = k \right\} \nonumber\\
& = \sum_{k}\Pr\left\{k^\star= k \mid \mathbf{I}(k^\star)=1 \right\}\Pr\left\{\tS^n[k]=\ts^n, S^n=s^n  \mid \mathbf{I}(k)=1, k^\star = k \right\} \nonumber\\
&\overset{a} =  \sum_{k}\Pr\left\{k^\star= k \mid\mathbf{I}(k^\star)=1 \right\}\Pr\left\{\tS^n[k]=\ts^n, S^n=s^n  \mid \mathbf{I}(k)=1 \right\} \nonumber\\
&= \sum_{k}\Pr\left\{k^\star= k \mid \mathbf{I}(k^\star)=1 \right\} \frac{\Pr \left\{\mathbf{I}(k)=1 \mid \tS^n[k] =\ts^n, S^n =s^n\right\}\Pr \left\{\tS^n[k] =\ts^n, S^n =s^n\right\}}{\Pr\left\{ \mathbf{I}(k)=1\right\}} \nonumber\\
& \overset{b} \leq \sum_{k}\Pr\left\{k^\star= k \mid \mathbf{I}(k^\star)=1\right\}  \frac{\frac{2^{-n(\overline{\mathbf{I}} [\bf{S};\bf{\tilde{S}}]+\gamma)}p_{\tS^nS^n(\ts^n,s^n)}}{p_{\tS^n}(\ts^n)p_{S^n}(s^n)}p_{\tS^n}(\ts^n) p_S^n(s^n)}{\left(1- \eps -\sqrt{\eps} -\eps^{\frac{1}{4}}\right)2^{-n(\overline{\mathbf{I}} [\bf{S};\bf{\tilde{S}}]+\gamma)}}\nonumber\\
\label{crit}
& = \frac{p_{\tS^n S^n}(\ts^n, s^n)}{\left(1- \eps -\sqrt{\eps} -\eps^{\frac{1}{4}}\right)},
\end{align}
where $a$ follows because $(\tS^n[k], S^n)$ is conditionally independent of $k^\star$ given the indicator random variable $\mathbf{I}(k)$ and $b$ follows from the definition of $\mathbf{I}(k)$ and from the fact that $\tS^n[k]$ is independent of $S^n$ and because of the following set of inequalities:
\begin{align*}
\Pr\left\{\mathbf{I}(k) =1\right\} &\geq \Pr\left\{\zeta(k) =1\right\} - \Pr \left\{\zeta(k) = 1, g_{1}(\tS^n[k]) \geq \sqrt{\eps} \right\} - \Pr \left\{\zeta(k) = 1, g_{2}(\tS^n[k])\geq {\eps^{\frac{1}{4}}} \right\} \nonumber\\
&\overset{a} \geq \sum_{(\ts^n,s^n) \in \cT_n(p_{\tS^nS^n})} \Pr\left\{\tS^n[k]=\ts^n\right\}\Pr \left\{S^n = s^n\right\}\frac{p_{S^n\tS^n}(s^n,\ts^n)}{2^{n(\overline{\mathbf{I}} [\bf{S};\bf{\tilde{S}}]+\gamma)}p_{\tS^n}(\ts^n)p_{S^n}(s^n)}\\
& \hspace{8mm}- \sum_{s^n, \ts^n : g_{1}(\ts^n) \geq  \sqrt{\eps}}\Pr\left\{\tS^n[k]=\ts^n\right\}\Pr \left\{S^n = s^n\right\}\frac{p_{S^n\tS^n}(s^n,\ts^n)}{2^{n(\overline{\mathbf{I}} [\bf{S};\bf{\tilde{S}}]+\gamma)}p_{\tS^n}(\ts^n)p_{S^n}(s^n)}\\
& \hspace{8mm}- \sum_{s^n,\ts^n : g_{2}(\ts^n) \geq \eps^{\frac{1}{4}}}\Pr\left\{\tS^n[k]=\ts^n\right\}\Pr \left\{S^n = s^n\right\}\frac{p_{S^n\tS^n}(s^n,\ts^n)}{2^{n(\overline{\mathbf{I}} [\bf{S};\bf{\tilde{S}}]+\gamma)}p_{\tS^n}(\ts^n)p_{S^n}(s^n)}\\
&= 2^{-n(\overline{\mathbf{I}} [\bf{S};\bf{\tilde{S}}]+\gamma)}\left(\Pr \left\{ \cT_n(p_{\tS^nS^n})\right\} - \Pr \left\{g_{1}(\tS^n) \geq \sqrt{\eps}\right\} -\Pr \left\{g_{2}(\tS^n) \geq {\eps}^{\frac{1}{4}}\right\} \right)\\
& \overset{b} \geq 2^{-n(\overline{\mathbf{I}} [\bf{S};\bf{\tilde{S}}]+\gamma)}\left(1- \eps -\sqrt{\eps} -\eps^{\frac{1}{4}}\right),
%& \overset{b}\geq (1-\eps)2^{-n(\overline{\mathbf{I}} [\bf{S};\bf{\tilde{S}}]+\gamma)},
\end{align*}
where $a$ follows from the definition of $\mathbf{I}(k)$ and $b$ follows from the definition of the set $\cT_n(p_{\tS^nS^n})$ and under the assumption that $n$ is large enough such that $\Pr\left\{\cT_n(p_{\tS^nS^n})\right\} \geq 1- \eps$ and from \eqref{g1m} and \eqref{g2m}. Combining \eqref{ne} and \eqref{crit} we now bound $2\mathbb{E}_{\cC_n^{(A)}\cC_n^{(C)} X^n S^n}\left[\mathbf{I}\{\cE_1^c\}\mathbf{I}\left\{\cE^c_2\right\} \tr\left[\left(\mathbb{I}-\Lambda_{U^n[\ell^\star]}\right) \rho^{B^n}_{X^n(U^n[\ell^\star],\tilde{S}^n[k^{\star}]),S^n}\right] \right]$ as follows:
\begin{align}
&2\mathbb{E}_{\cC_n^{(A)}\cC_n^{(C)} X^n S^n}\left[\mathbf{I}\{\cE_1^c\}\mathbf{I}\left\{\cE^c_2\right\} \tr\left[\left(\mathbb{I}-\Lambda_{U^n[\ell^\star]}\right) \rho^{B^n}_{X^n(U^n[\ell^\star],\tilde{S}^n[k^{\star}]),S^n}\right] \right] \nonumber\\
&\hspace{-1mm} \overset{a}\leq2\sum_{\ts^n,s^n,u^n,x^n} \Pr\left\{\tS^n[k^\star]=\ts^n, S^n=s^n\mid \mathbf{I}(k^\star)=1\right\}\Pr \bigg\{ U^n[\ell^\star]=u^n\mid S^n =s^n, \tS^n[k^\star]= \ts^n, \nonumber\\
&\hspace{30mm}\mathbf{I}(k^\star) =1, \mathbf{J}(k^\star,\ell^\star)=1\bigg\}
\Pr\left\{X^n = x^n \mid U^n[\ell^\star]=u^n, \tS^n[k^\star]=\ts^n, S^n= s^n\right\} \nonumber\\
&\hspace{22mm} \tr\left[\left(\mathbb{I}-\Lambda_{(U^n[\ell^\star]=u^n)}\right) \rho^{B^n}_{X^n(U^n[\ell^\star]=u^n,\tS^n[k^\star] = \tilde{s}^n) =x^n, S^n = s^n}\right] \nonumber \\
& \hspace{-1mm}\overset{b} \leq \frac{2}{\left(1- \eps -\sqrt{\eps} -\eps^{\frac{1}{4}}\right)}\sum_{\ts^n,u^n} p_{\tS^n}(\ts^n)\Pr \bigg\{ U^n[\ell^\star]=u^n\mid \tS^n[k^\star]= \ts^n, \mathbf{J}(k^\star,\ell^\star)=1\bigg\}\nonumber\\
&\hspace{10mm} \tr\left[\left(\mathbb{I}-\Lambda_{(U^n[\ell^\star]=u^n)}\right) \sum_{x^n, s^n}p_{S^n \mid \tS^n} (s^n \mid \ts^n)p_{X^n \mid U^n\tS^n}(x^n \mid u^n,\ts^n)\rho^{B^n}_{X^n(U^n[\ell^\star]=u^n,\tS^n[k^\star] = \tilde{s}^n) =x^n, S^n = s^n}\right] \nonumber\\
&\hspace{-1mm} \overset{c}= \frac{2}{\left(1- \eps -\sqrt{\eps} -\eps^{\frac{1}{4}}\right)}\sum_{\ts^n,u^n} p_{\tS^n}(\ts^n)\Pr \bigg\{ U^n[\ell^\star]=u^n\mid \tS^n[k^\star]= \tS^n, \mathbf{J}(k^\star,\ell^\star)=1\bigg\} \nonumber\\
& \hspace{45mm}  \tr\left[\left(\mathbb{I}-\Lambda_{(U^n[\ell^\star]=u^n)}\right) \rho^{B^n}_{(U^n[\ell^\star]=u^n,\tS^n[k^\star] = \tilde{s}^n)}\right] \nonumber\\
& \hspace{-1mm} \overset{d} \leq  \frac{2\sqrt{\eps}}{\left(1- \eps -\sqrt{\eps} -\eps^{\frac{1}{4}}\right)} \sum_{\ts^n,u^n} p_{\tS^n}(\ts^n)\Pr \bigg\{ U^n[\ell^\star]=u^n\mid \tS^n[k^\star]= \ts^n, \mathbf{J}(k^\star,\ell^\star)=1\bigg\}\nonumber\\
\label {333}
&\hspace{-1mm}  = \frac{2\sqrt{\eps}}{\left(1- \eps -\sqrt{\eps} -\eps^{\frac{1}{4}}\right)},
\end{align}
where $a$ follows from \eqref{ne}; $b$ follows from \eqref{crit} and from the fact that given $(U^n[\ell^\star]=u^n, \tS^n[k^\star]=\ts^n)$, $(k^\star,\ell^\star)$ is deterministic, thus, $X^n \mid \left\{U^n[\ell^\star]=u^n, \tS^n[k^\star]=\ts^n, S^n= s^n\right\} \sim p_{X^n \mid U^n\tS^nS^n}(x^n \mid u^n, \ts^n,s^n) = p_{X^n \mid U^n\tS^n}(x^n \mid u^n, \ts^n)$ and $U^n[\ell^\star]$ is conditionally independent of $(S^n, \mathbf{I}(k^\star))$ given $(\tS^n[k^\star]=\ts^n, \mathbf{J}(k^\star,\ell^\star)=1 )$; $c$ follows because from the definition of $\rho^{B^n}_{u^n\ts^n}$ mentioned in \eqref{rhofus} and $d$ follows because for $\mathbf{J}(k^\star,\ell^\star) =1$, we have $ \tr\left[\left(\mathbb{I}-\Lambda_{(U^n[\ell^\star]=u^n)}\right) \rho^{B^n}_{(U^n[\ell^\star]=u^n,\tS^n[k^\star] = \tilde{s}^n)}\right]< \sqrt{\eps}.$

{{$4 \sum_{\ell^\prime \neq \ell^\star}\mathbb{E}_{\cC_n^{(A)}\cC_n^{(C)}X^nS^n}\left[\mathbf{I}\left\{\cE_1^c\right\}\mathbf{I}\left\{\cE^c_2\right\} \tr\left[\Lambda_{U^n(\ell^\prime)} \rho^{B^n}_{X^n(U^n[\ell^\star],\tilde{S}^n[k^{\star}]),S^n}\right] \right]$ is bounded as follows:}}
\setlength{\belowdisplayskip}{0pt}%\begin{align}
  \begin{alignat}{3}
&4 \sum_{\ell^\prime \neq \ell^\star}\mathbb{E}_{\cC_n^{(A)}\cC_n^{(C)} X^n S^n}\left[\mathbf{I}\left\{\cE_1^c\right\}\mathbf{I}\left\{\cE^c_2\right\} \tr\left[\Lambda_{U^n[\ell^\prime]} \rho^{B^n}_{X^n(U^n[\ell^\star],\tilde{S}^n[k^{\star}]),S^n}\right] \right]\nonumber\\
&\hspace{-1mm}\overset{a} =4 \sum_{\ell^\prime \neq \ell^\star}\mathbb{E}_{\cC_n^{(A)}\cC_n^{(C)} X^n S^n}\sum_{k,\ell}\mathbf{I}\left\{k^\star = k\right\}\mathbf{I}\left\{k^\star=k, \ell^\star = \ell\right\}\tr\left[\Lambda_{U^n[\ell^\prime]} \rho^{B^n}_{X^n(U^n[\ell^\star],\tilde{S}^n[k^{\star}]),S^n}\right] \nonumber\\
 &\hspace{-1mm}\overset{b}\leq4 \sum_{k, \ell, \ell^\prime \neq \ell}\mathbb{E}_{\cC_n^{(A)}\cC_n^{(C)} X^n S^n}\mathbf{I}(k)\mathbf{J}(k,\ell)\tr\left[\Lambda_{U^n[\ell^\prime]} \rho^{B^n}_{X^n(U^n[\ell],\tilde{S}^n[k]),S^n}\right] \nonumber\\
 &\hspace{-1mm} =4 \sum_{k,\ell^\prime \neq \ell}\sum_{u^{\prime n}, u^n,\ts^n,x^n,s^n} \Pr\bigg\{U^n[l^\prime] = u^{\prime n}, U^n[\ell] = u^n, S^n=s^n, \tS^n[k] =\ts^n, X^n(U^n[\ell],\tS^n[k]) = x^n, \nonumber\\
 &\hspace{45mm} \mathbf{I}(k)=1, \mathbf{J}(k,\ell)=1\bigg\} \tr\left[\Lambda_{u^{\prime n}} \rho^{B^n}_{X^n(U^n[\ell]=u^n,\tS^n[k] = \tilde{s}^n) =x^n, S^n = s^n}\right] \nonumber\\
&\hspace{-1mm} =4 \sum_{k, \ell^\prime \neq \ell}\sum_{u^{\prime n}, u^n,\ts^n, x^n, s^n}\hspace{-6mm} p_{U^n}(u^{\prime n})p_{U^n}(u^n)p_{\tilde{S}^n}(\tilde{s}^n)p_{X^n \mid U^n\tS^n}(x^n \mid u^n,\ts^n)p_{S^n}(s^n)\Pr\left\{\mathbf{I}(k)=1 \mid S^n =s^n, \tilde{S}^n[k] = \tilde{s}^n\right\} \nonumber\\
&\hspace{30mm}\Pr\left\{\mathbf{J}(k, \ell)=1 \mid \mathbf{I}(k)=1, S^n =s^n, U^n[\ell] = u^n,\tilde{S}^n[k] = \tilde{s}^n\right\} \nonumber\\
&\hspace{30mm}\tr\left[\Lambda_{u^{\prime n}} \rho^{B^n}_{X^n(U^n[\ell]=u^n,\tS^n[k] = \tilde{s}^n) =x^n, S^n = s^n}\right] \nonumber\\
 &\hspace{-1mm}\overset{c}\leq4 \sum_{k, \ell, \ell^\prime \neq \ell} \sum_{u^{\prime n}, u^n,\ts^n, x^n, s^n} p_{U^n}(u^{\prime n})p_{U^n}(u^n)p_{\tilde{S}^n}(\tilde{s}^n)p_{X^n \mid U^n\tS^n}(x^n \mid u^n,\ts^n)p_{S^n}(s^n)2^{-n(\overline{\mathbf{I}}[\bf{S};\bf{\tilde{S}}]+\gamma)}\frac{p_{S^n\tilde{S}^n}(s^n,\tilde{s}^n)}{p_{S^n}(s^n)p_{\tilde{S}^n}(\tilde{s}^n)} \nonumber\\
&\hspace{35mm} 2^{-n(\overline{\mathbf{I}}[\bf{U};\bf{\tilde{S}}]+\gamma)}\frac{p_{U^n\tilde{S}^n}(u^n,\tilde{s}^n)}{p_{U^n}(u^n)p_{\tilde{S}^n}(\tilde{s}^n)}
 \tr\left[\Lambda_{u^{\prime n}} \rho^{B^n}_{X^n(U^n[\ell]=u^n,\tS^n[k] = \tilde{s}^n) =x^n, S^n = s^n}\right] \nonumber\\%\tr
 &\hspace{-1mm} =4  \times 2^{-n\left(\overline{\mathbf{I}}[\bf{U};\bf{\tilde{S}}] + \overline{\mathbf{I}}[\bf{S};\bf{\tilde{S}}] +2\gamma\right)}\sum_{k, \ell, \ell^\prime \neq \ell} \sum_{u^n,\ts^n}p_{U^n\tS^n}(u^n\ts^n)p_{X^n \mid U^n\tS^n}(x^n \mid u^n,\ts^n)\sum_{u^{\prime n}}p_{U^n}(u^{\prime n}) \nonumber\\
& \hspace{45mm} \tr\left[\Lambda_{u^{\prime n}}\sum_{s^n,x^n}p_{S^n \mid \tS^n}(s^n \mid \ts^n)p_{X^n \mid U^n\tS^n}(x^n \mid u^n,\ts^n) \rho^{B^n}_{X^n(U^n[\ell]=u^n,\tS^n[k] = \tilde{s}^n) =x^n, S^n = s^n}\right] \nonumber\\
&\hspace{-1mm}\overset{d}= 4  \times 2^{-n\left(\overline{\mathbf{I}}[\bf{U};\bf{\tilde{S}}] + \overline{\mathbf{I}}[\bf{S};\bf{\tilde{S}}] + 2\gamma \right)}\sum_{k, \ell, \ell^\prime \neq \ell} \sum_{u^n,\ts^n}p_{U^n\tS^n}(u^n\ts^n)
\sum_{u^{\prime n}}p_{U^n}(u^{\prime n})\tr\left[\Lambda_{u^{\prime n}} \rho^{B^n}_{u^n, \ts^n}\right] \nonumber\\
&\hspace{-1mm}\overset{e}= 4  \times 2^{-n\left(\overline{\mathbf{I}}[\bf{U};\bf{\tilde{S}}] + \overline{\mathbf{I}}[\bf{S};\bf{\tilde{S}}] + 2 \gamma \right)}\sum_{k, \ell, \ell^\prime \neq \ell} \sum_{u^n,\ts^n}p_{U^n\tS^n}(u^n\ts^n)\sum_{u^{\prime n}}p_{U^n}(u^{\prime n})\tr\left[\tr_{U^n}\left[\Pi^{U^nB^n}\left(\ket{u^{\prime n}}\bra{u^{\prime n}}\otimes \mathbb{I}\right)\right] \rho^{B^n}_{u^n, \ts^n}\right] \nonumber\\
&\hspace{-1mm}= 4  \times 2^{-n\left(\overline{\mathbf{I}}[\bf{U};\bf{\tilde{S}}] + \overline{\mathbf{I}}[\bf{S};\bf{\tilde{S}}] + 2\gamma \right)}\sum_{k, \ell, \ell^\prime \neq \ell} \sum_{u^n,\ts^n}p_{U^n\tS^n}(u^n\ts^n)\sum_{u^{\prime n}}p_{U^n}(u^{\prime n})\tr\left[\tr_{U^n}\left[\Pi^{U^nB^n}\left(\ket{u^{\prime n}}\bra{u^{\prime n}}\otimes \rho^{B^n}_{u^n, \ts^n}\right)\right] \right] \nonumber\\
&\hspace{-1mm}= 4  \times 2^{-n\left(\overline{\mathbf{I}}[\bf{U};\bf{\tilde{S}}] + \overline{\mathbf{I}}[\bf{S};\bf{\tilde{S}}]+2\gamma\right)}\nonumber\\
&\hspace{3mm}\sum_{k, \ell, \ell^\prime \neq \ell} \tr\bigg[\tr_{U^n}\bigg[\Pi^{U^nB^n}\bigg(\sum_{u^{\prime n}}p_{U^n}(u^{\prime n})\ket{u^{\prime n}}\bra{u^{\prime n}}\otimes \sum_{(u^n,\ts^n)}p_{U^n}(u^n)p_{\tS^n \mid U^n}(\ts^n \mid u^n)\rho^{B^n}_{u^n, \ts^n}\bigg)\bigg] \bigg] \nonumber\\
&\hspace{-1mm} = 4  \times 2^{-n\left(\overline{\mathbf{I}}[\bf{U};\bf{\tilde{S}}] + \overline{\mathbf{I}}[\bf{S};\bf{\tilde{S}}] +2\gamma\right)}\sum_{k, \ell, \ell^\prime \neq \ell} \tr\left[\Pi^{U^nB^n}\left(\Theta^{U^n }\otimes \Theta^{B^n}\right)\right]  \nonumber\\
&\hspace{-1mm}\overset{f}\leq 4 \times 2^{-n\left(\overline{{\mathbf{I}}}[\bf{U};\bf{\tilde{S}}] + \overline{{\mathbf{I}}}[\bf{S};\bf{\tilde{S}}] + \underline{{\mathbf{I}}}[{\bf{U}};{\bf{B}}] + \gamma\right) }2^{n(R+2r+R_S)} \nonumber\\
&\hspace{-1mm} \overset{g} = 4 \times 2^{-n\gamma}, \nonumber\\
\label{444}
& \hspace{-1mm} \overset{h}  \leq 4 \eps,
\end{alignat}
 %\end{align}
 where $a$ follows from the union bound; $b$ follows because $\mathbf{I}\left\{k^\star = k\right\} =\mathbf{I}\left\{k^\star = k\right\}\mathbf{I}(k) \leq \mathbf{I(k)}$ and $\mathbf{I}\left\{k^\star = k, \ell^\star = \ell \right\} =\mathbf{I}\left\{k^\star = k, \ell^ \star= \ell \right\}\mathbf{J}(k,l) \leq \mathbf{J}(k,\ell)$, $c$ follows from the definition of $\mathbf{I}(k)$ and $\mathbf{J}(k,\ell)$, $d$ follows from the definition of $\rho^{B^n}_{u^n,\ts^n}$ mentioned in \eqref{rhofus}, $e$ follows from the definition of $\Lambda_{u^{\prime n}}$
mentioned in \eqref{lambda}, $f$ follows from the definition of $\Pi^{U^nB^n}$ mentioned in \eqref{pi} and because $k$ ranges over $[1:2^{nR_S}],$ $\ell$ ranges over $[1:2^{nr}]$ and $\ell^\prime$ ranges over $[1:2^{n(R+r)}]$ and $g$ follows because of our choice of $R,r$ and $R_S$ and $h$ follows under the assumption that $n$ is large enough such that $2^{-n\gamma} \leq \eps.$ Thus, it now follows from \eqref{errorsideinf}, \eqref{e1}, \eqref{n1}, \eqref{firster}, \eqref{333} and \eqref{444} that
\beq 
\Pr\left\{\tilde{m} \neq 1\right\} \leq 6\eps+3\sqrt{\eps} + 3\eps^{\frac{1}{4}}+\frac{2\sqrt{\eps}}{\left(1- \eps -\sqrt{\eps} -\eps^{\frac{1}{4}}\right)}+3\exp(-2^{n\gamma}). \nonumber
\enq
This completes the proof for achievability.
\subsection{Converse}
Suppose $(R,R_S)$ be an achievable rate pair. It then follows from Definition \ref{ach} that there exists an $(n,M_n,M_{e,n}, \eps_n)$ code such that $R \leq \liminf_{n\to \infty} \frac{\log M_n}{n}$ and $R_S \geq \limsup_{n\to \infty} \frac{\log M_{e,n}}{n}.$ Let
\beq \label {conts}\tS^n = f_{e,n}(S^n),\enq where $f_{e,n}$ is defined in Definition \ref{code}. Also, let $U^n$ represent an arbitrary random variable denoting the uniform choice of a message in $[1:M_n]$.
Notice that the message random variable is independent of $S^n.$ Hence, from the definition of $U^n$ and $\tS^n$ it now follows that $U^n$ and $\tS^n$ are independent of each other. Thus, \beq\label{ens}\overline{\mathbf{I}}[{\bf{U}};{\bf{\tS}}] = 0.\enq
Further, notice that in the setting of the problem and for the choice of $U^n$ and $\tS^n$ fixed above the following classical-quantum state is induced
\begin{align}
&\sigma^{S^nU^nX^nB^n}= \sum_{(s^n,\ts^n,u^n,x^n)}p_{S^n}(s^n)p_{\tS^n \mid S^n}(\ts^n \mid s^n)p_{U^n} (u^n)p_{X^n \mid \tS^n U^n}(x^n \mid \ts^n,u^n)\ket{s^n}\bra{s^n}^{S^n}  \nonumber \\
&\hspace{45mm}\otimes \ket{\ts^n}\bra{\ts^n}^{\tS^n} \otimes \ket{u^n}\bra{u^n}^{U^n}\otimes \ket{x^n}\bra{x^n}^{X^n} \otimes \rho^{B^n}_{x^n,s^n} .
\end{align}
We will now first prove the lower bound on $R_S$. Towards this notice that from \eqref{conts} and from the definition of $f_{e,n}$ it follows that the cardinality of the set over which the random variable $\tS^n$ takes values cannot be larger than $M_{e,n}$. Thus, it now follows that from \cite[Lemma 2.6.2]{han-book} that 
\beq
\label{con1}
\Pr\left\{\frac{1}{n}\log \frac{1}{p_{\tS^n}} \geq \frac{1}{n} \log M_{e,n}+ \gamma \right\} \leq 2^{-n\gamma},
\enq
where $\gamma >0$ is an arbitrary constant. Furthermore, since $\frac{1}{n}\log \frac{p_{\tS^n \mid S^n}}{p_{\tS^n}} \leq \frac{1}{n}\log \frac{1}{p_{\tS^n}}$ it now follows from \eqref{con1} that 
 \beq
 \label{conv2}
 \Pr\left\{\frac{1}{n}\log \frac{p_{\tS^n \mid S^n}}{p_{\tS^n}} \geq \frac{1}{n} \log M_{e,n}+ \gamma \right\} \leq 2^{-n\gamma}.
 \enq
Thus, from the Definition \ref{limsup} and \eqref{conv2} it now follows that there exists an $n_0$ such that for every $n >n_o,$ we have
\begin{align*}
\overline{\mathbf{I}} [\mathbf{S};\mathbf{\tS}] &\leq \frac{1}{n} \log M_{e,n} + \gamma \\
& \leq \limsup _{n \to  \infty}\frac{1}{n} \log M_{e,n} + 2\gamma\\
& \leq R_S + 2\gamma,
\end{align*}
where the last inequality follows from the definition of $R_S.$

We now prove the upper bound on $R.$   Let $\rho^{B^n}_{u^n}= \sum_{(s^n,\ts^n,x^n)}p_{S^n }(s^n)p_{\tS^n \mid S^n}(\ts^n \mid s^n)p_{X^n \mid U^n\tS^n}(x^n \mid u^n,s^n) \rho^{B^n}_{x^n,s^n}$ and $\rho^{B^n} = \mathbb{E}\left[\rho^{B_n}_{U^n}\right],$ where $U^n$ as mentioned above is uniformly distributed over the set $[1:M_n]$ and $\tS^n$ is as defined in \eqref{conts}. Fix $\gamma>0$. It now follows from Definition \ref{code} and \cite[Lemma $4$]{hayshi-nagaoka-2002} that $\eps_n$ satisfies the following bound,
 \beq
\label{conerr}
\eps_n \geq \sum_{u^n\in \cU^n} p_{U^n}(u^n)\tr \left[\rho^{B^n}_{u^n} \left\{\rho^{B^n}_{u^n} \preceq 2^{n\left(\frac{1}{n}\log M_n -\gamma\right)} \rho^{B^n}\right\}\right] - 2^{-n\gamma}.
\enq
Letting $\frac{1}{n}\log M_n = \underline{{\mathbf{I}}}[{\bf{U}};{\bf{B}}]+ 2\gamma$ in \eqref{conerr} it holds that 
\beq
\eps_n \geq \sum_{u^n\in \cU^n} p_{U^n}(u^n)\tr \left[\rho^{B^n}_{u^n} \left\{\rho^{B^n}_{u^n} \preceq 2^{n\left(\underline{{\mathbf{I}}}[{\bf{U}};{\bf{B}}]+ \gamma\right)} \rho^{B^n}\right\}\right] - 2^{-n\gamma}.
\enq
However, while $2^{-n\gamma} \to 0$ as $n \to \infty$, the definition of $\underline{{\mathbf{I}}}[{\bf{U}};{\bf{B}}]$ implies the existence of $\eps_o > 0$ and infinitely many $n'$s satisfying
$\sum_{u^n\in \cU^n} p_{U^n}(u^n)\tr \left[\rho^{B^n}_{u^n} \left\{\rho^{B^n}_{u^n} \prec 2^{n\left(\frac{1}{n}\log M_n -\gamma\right)} \rho^{B^n}\right\}\right]> \eps_o.$ This further implies that $R \leq \liminf_{n\to \infty} \frac{\log M_n}{n}$ must satisfy the following for arbitrarily small probability of error:
\begin{align}
R & \leq \underline{{\mathbf{I}}}[{\bf{U}};{\bf{B}}] + 2 \gamma \nonumber\\
 & \leq \underline{{\mathbf{I}}}[{\bf{U}};{\bf{B}}] - \overline{{\mathbf{I}}}[{\bf{U}};{\bf{\tS}}] + 2 \gamma \nonumber\\
 & \leq \sup_{\{\omega_n\}_{n=1}^{\infty}}\left(\underline{{\mathbf{I}}}[{\bf{U}};{\bf{B}}] - \overline{{\mathbf{I}}}[{\bf{U}};{\bf{\tS}}]\right) + 2\gamma,\nonumber
\end{align} 
where the second inequality follows from \eqref{ens} and in the last inequality the supremum is taken over all  sequence of classical-quantum states having the following form for every $n$,
 \begin{align*}
 \omega_n& := \sum_{(s^n,\ts^n,u^n,x^n)}p_{S^n}(s^n)p_{\tS^n \mid S^n}(\ts^n \mid s^n)p_{U^n \mid \tS^n} (u^n \mid \ts^n)p_{X^n \mid \tS^n U^n}(x^n \mid \ts^n,u^n)\ket{s^n}\bra{s^n}^{S^n} \otimes \ket{\ts^n}\bra{\ts^n}^{\tS^n}  \nonumber \\
&\hspace{25mm}\otimes \ket{u^n}\bra{u^n}^{U^n}\otimes \ket{x^n}\bra{x^n}^{X^n} \otimes \rho^{B^n}_{x^n,s^n} .
\end{align*}
This completes the proof for the converse.

\section{Conclusion and Acknowledgement} 
We extended the result of Heggard and El Gamal \cite{heegard-gamal-1983} to the quantum case in the information-spectrum setting. The proof in \cite{heegard-gamal-1983} is based on the covering lemma \cite[Lemma 3.3]{Gamal-Kim-book}, conditional typicality lemma \cite{Gamal-Kim-book} and Markov lemma \cite[Lemma 12.1]{Gamal-Kim-book}. We have shown in this paper that quantum information-spectrum generalization of the result of Heggard and El Gamal \cite{heegard-gamal-1983} can be derived even in the absence of these powerful lemmas.  A natural open question which arises from the problem studied in this manuscript is to study the problem of communication when side information is available at the decoder end. This study along with the techniques presented in this paper will then lead to the establishing of the capacity region when the side information is available both at the encoder and decoder end. 

This work was supported by the Engineering and Physical Sciences Research Council (Grant No. EP/M013243/1).

\bibliographystyle{ieeetr}
\bibliography{master}

\begin{thebibliography}{10}

\bibitem{wilde-book}
{M. M. Wilde}, ``From classical to quantum {Shannon} theory.''
  http://arxiv.org/abs/1106.1445, 2011.

\bibitem{covertom}
{T. M. Cover} and {J. A. Thomas}, {\em Elements of {I}nformation {T}heory}.
\newblock Hoboken, NJ, USA: Wiley, 2nd~ed., 2006.

\bibitem{gray-book}
{R. M. Gray}, {\em Entropy and Information Theory}.
\newblock New York, USA: Springer-Verlag, 1990.

\bibitem{han-verdu-spectrum-94}
{S. Verd\'{u}} and {T. S. Han}, ``A general formula for channel capacity,''
  {\em IEEE Trans. Inf. Theory}, vol.~40, pp.~1147--1157, 1994.

\bibitem{miyakaye-kanaya-1995}
{S. Miyake} and {F. Kanaya}, ``Coding theorems on correlated general sources,''
  {\em IEICE Trans. Fundamentals}, vol.~E78-A(9), pp.~1063--1070, Sept. 1995.

\bibitem{muramatsu}
{K.- I. Iwata} and {J. Muramatsu}, ``An information-spectrum approach to
  rate-distortion function with side information,'' {\em IEICE Trans. on
  Fundamentals of Electronics, Communication and Computers}, vol.~E85-A,
  pp.~1387--1395, June 2002.

\bibitem{hayashi-classical-non-iid}
{M. Hayashi}, ``General nonasymptotic and asymptotic formulas in channel
  resolvability and identification capacity and their application to the
  wiretap channel,'' {\em IEEE Trans. Inf. Theory}, vol.~52, pp.~1562--1575,
  April 2006.

\bibitem{arbitrary-wiretap}
{M. Bloch} and {J. N. Laneman}, ``On the secrecy capacity of arbitrary wiretap
  channel,'' in {\em Proc. Allerton Conf. Commun. Control, Computing},
  (Monticello, IL, USA), Sept. 2008.

\bibitem{hayshi-nagaoka-2002}
{H. Nagaoka} and {M. Hayashi}, ``An information-spectrum approach to classical
  and quantum hypothesis testing for simple hypotheses,'' {\em IEEE Trans. Inf.
  Theory}, vol.~53, pp.~534--549, Feb. 2007.

\bibitem{ogawa-nagaoka-2000-strong}
{T. Ogawa} and {H. Nagaoka}, ``Strong converse and {S}tein's lemma in quantum
  hypothesis testing,'' {\em IEEE Trans. Inf. Theory}, vol.~46, pp.~2428--2433,
  Nov. 2000.

\bibitem{Hayashi-noniid}
{M. Hayashi} and {H. Nagaoka}, ``General formulas for capacity of
  clasical-quantum channels,'' {\em IEEE Trans. Inf. Theory}, vol.~49,
  pp.~1753--1768, 2003.

\bibitem{hayashi-entanglement-2006}
{M.Hayashi}, ``General formulas for fixed-length entanglement concentration,''
  {\em IEEE Trans. Inf. Theory}, vol.~49, pp.~1753--1768, May 2003.

\bibitem{datta-byeondiid-2006}
{G. Bowen} and {N. Datta}, ``Beyon i.i.d. in quantum information theory.''
  http://arxiv.org/abs/quant-ph/0604013, Oct. 2006.

\bibitem{datta-renner-2009}
{N. Datta} and {R. Renner}, ``Smooth {R}\'{e}nyi entropies and the quantum
  information spectrum,'' {\em IEEE Trans. Inf. Theory}, vol.~55,
  pp.~2807--2815, 2009.

\bibitem{radhakrishnan-sen-warsi-archive}
{J. Radhakrishnan}, {P. Sen}, and {N. Warsi}, ``One-shot marton inner bound for
  classical-quantum broadcast channel.'' http://arxiv.org/abs/1410.3248, Oct.
  2014.

\bibitem{heegard-gamal-1983}
{C. Heggard} and {A. El Gamal}, ``On the capacity of computer memory with
  defects,'' {\em IEEE Trans. Inf. Theory}, vol.~29, pp.~731--739, Sept. 1983.

\bibitem{gelfand-pinsker}
{S. Gel'fand} and {M. Pinsker}, ``Coding for channel with random parameters,''
  {\em Prob.of Control and Inf. Th.}, vol.~9, no.~1, pp.~19--31, 1980.

\bibitem{shannon1948}
{C. E. Shannon}, ``A mathematical theory of communication,'' {\em Bell Sys.
  Tech. J.}, vol.~27, pp.~379--423 and 623--656, July and {O}ct. 1948.

\bibitem{wang-renner-prl}
{L. Wang} and {R. Renner}, ``One-shot classical-quantum capacity and hypothesis
  testing,'' {\em Phys. Rev. Lett.}, vol.~108, pp.~200501--200505, May 2012.

\bibitem{han-book}
{T. S. Han}, {\em Information-Spectrum Methods in Information Theory}.
\newblock Berlin, Germany: Springer-Verlag, 2003.

\bibitem{Gamal-Kim-book}
{A. El Gamal} and {Y. H. Kim}, {\em Network Information Theory}.
\newblock Cambridge, U.K: Cambridge University Press, 2012.

\end{thebibliography}

\end{document}